\documentclass[doublecol]{epl2} 
\usepackage{graphicx}% Include figure files
\usepackage{dcolumn}% Align table columns on decimal point
\usepackage{bm}% bold math

\title{Ultrasolitons: multistability and subcritical power threshold\\ 
from higher-order Kerr terms}
\shorttitle{Ultrasolitons: multistability \& subcritical power threshold from HOKE} %Insert here a short version of the title if it exceeds 70 characters

\author{David Novoa\inst{1}\thanks{E-mail: \email{dnovoa@clpu.es}}\and Daniele Tommasini\inst{2} \and Humberto Michinel\inst{2}}
\shortauthor{D. Novoa \etal}

\institute{                    
  \inst{1} Centro de L\'aseres Pulsados Ultracortos Ultraintensos (CLPU) - Plaza de la Merced s/n, Salamanca, ES-37008 Spain.\\
  \inst{2}  Departamento de F\'{\i}sica Aplicada. Universidade de Vigo - As Lagoas s/n, Ourense, ES-32004 Spain.}
  

\abstract{
We show that an optical system involving competing higher-order Kerr nonlinearities 
can support the existence of \emph{ultrasolitons}, namely extremely localized modes 
that only appear above a certain threshold for the central intensity. 
Such new solitary waves can be produced for powers \emph{below} the 
usual collapse threshold, but they can also coexist with ordinary, 
lower-intensity solitons. We derive analytical conditions for the occurrence 
of multistability and analyze the dynamics of the different kinds of fundamental 
eigenmodes that can be excited in these nonlinear systems. 
We also discuss the possible transitions between solitary waves belonging 
to different nonlinear regimes through the mechanism of soliton switching.}

\begin{document}

\maketitle

\section{Introduction}

%---------------- INTRODUCTION --------------------

A recent measurement of the instantaneous higher-order Kerr (HOKE) 
coefficients in gases\cite{loriot}  has led to a revolutionary description 
of the filamentation of ultrashort laser pulses\cite{bejot}. 
The filament stabilization is attributed to the competition of  the HOKE  
focusing and defocusing contributions to the refractive index alone, 
rather than to their interplay with the ionization-induced plasma defocusing, 
which had a key role in the traditional interpretation\cite{couairon}. 
This new paradigm has stimulated an increasing amount of 
work\cite{theory1,theory2,experim1,experim2,experim3,experim4,experim5,experim6}, 
aimed either at testing the controversial results reported in  Refs. \cite{loriot,bejot}, 
or at the theoretical exploration of its rich phenomenological 
implications\cite{fermionic,kolesik1,kasparian1,kasparian2,stegeman,wang}.

In this Letter, we will demonstrate that, within a well-defined parameters region, 
a system involving just local HOKE nonlinearities can support 
the existence of a new branch of localized solutions 
that will be called \emph{ultrasolitons}. 
Such stationary states coexist with solitons of lower intensity, 
similar to those found for common optical media\cite{fermionic}. 
Very remarkably, this implies the emergence of \emph{optical soliton multistability} (OSM), 
i.e. the existence of two or more stationary states with the same power and 
different propagation constants and profiles, like in the systems reported in Refs. \cite{kaplan,kaplan2,kivshar,matusz11} and in the recent work \cite{singular}, that has appeared while we were preparing the revised version of the present paper.
We will also derive an analytical condition on the HOKE coefficients for the emergence of OSM, 
and show that the ultrasolitons can be \emph{subcritical}, 
i.e. they may exhibit powers even below the ordinary collapse threshold\cite{marburger}. 
Finally, we will discuss the transitions among multistable states through efficient 
\emph{soliton switching} processes.

\section{Mathematical model}

Let us consider a wave system evolving along the $\eta$ 
direction in the space of transverse coordinates $\xi$ and $\chi$, and assume
that the complex wavefunction $\Psi(\xi,\chi,\eta)$ satisfies 
the (dimensionless) nonlinear Schr\"odinger equation
\begin{equation}
  \label{NLSE} i \frac{\partial \Psi}{\partial \eta} + \frac{1}{2}
  \nabla_{\perp}^2 \Psi + F(\vert\Psi\vert^2) \Psi = 0,
\end{equation}
where
$\nabla_{\perp}^2 =\partial^2 / \partial \xi^2 + \partial^2 / \partial \chi^2$
and
\begin{equation}
  \label{index} F(\vert\Psi\vert^2) =\sum_{q}(-1)^{q+1} f_{2 q} \vert\Psi\vert^{2 q}.
\end{equation}
This formalism applies to the paraxial propagation 
of the linearly polarized electric field $E(x,y,z)$ of a laser pulse of mean wavenumber in
vacuum $k_0 = 2 \pi / \lambda_0$, being $\lambda_0$ the central wavelength, 
in a nonlinear optical medium whose refractive index depends upon the intensity 
$I=\epsilon_0 c E^2$ as
$n=n_0+\Delta n=n_0+ \sum_{q=1}^{4} n_{2 q} I^{q}$. In fact, motivated by the results of 
Ref. \cite{loriot} for the optical response of common gases, we will assume that 
the refractive index involves $4$ terms of increasing powers
in the intensity $I$ of the beam, 
with alternating sign coefficients $n_2, n_6 > 0$ and $n_4, n_8 < 0$, that
contribute to focusing and defocusing respectively. 
We will choose the dimensionless quantities such that $f_2=f_4=1$.
The relations between the dimensional and dimensionless physically-relevant quantities are then
$(\xi,\chi)=(n_0/\vert n_4\vert)^{1/2}(k_0n_2) (x,y)$, 
$\eta=(k_0n_2^2/\vert n_4\vert)z$, 
$\Psi(\xi,\chi,\eta)=(\epsilon_0 c \vert n_4\vert/n_2)^{1/2} E(x,y,z)$, 
$\Delta n=({n_2^2}/{\vert n_ 4\vert })F$, $n_6=({n_4^2}/{n_ 2})f_6$ and $n_8=({n_4^3}/{n_ 2^2})f_8$.
The only free parameters included in Eq. \ref{NLSE}  
are then $f_6$ and $f_8$,  that will be assumed to be positive.

For the sake of clarity, and motivated by Refs.\cite{bejot,loriot,fermionic},
in this Letter we will neglect the effects of multiphoton absorption, ionization 
and temporal dispersion of the optical pulses. 
However, we emphasize that our results may also be applied 
to multilevel atomic media, where the dependence of $\Delta n$ given by Eq. \ref{index}
may be achieved by the coherent control of the atomic ensemble 
via quantum-engineering techniques\cite{eitCQ}. 

\section{Analytical condition for multistability}

Let us now derive an analytical condition for the possible emergence of OSM. 
Assuming radial symmetry, we search for soliton solutions of the form 
$\Psi (\xi,\chi,\eta) = \Phi (r) e^{- i \mu \eta}$, where $r\equiv \sqrt{\xi^2 + \chi^2}$ and $\mu$ is the
propagation constant. As discussed in Refs. \cite{fermionic,dripping}, 
$\mu$ can be identified with the chemical potential of an
equivalent thermodynamical 2D system of $N = \int
\rho d\xi d\chi \equiv \int | \Phi |^2 d\xi d\chi$ particles 
(for the optical system, $N=n_0k_0^2 n_2 {\cal P}$, where ${\cal P}$ is the total power of the optical field). 
Eq. \ref{NLSE} can then be derived by minimizing the Landau's
grand potential{\cite{landau}} $\Omega = - \int pd\xi d\chi$, where the pressure
field $p$ is
\begin{equation}
  p = - \frac{1}{2} | \nabla_{\perp} \Phi |^2 + \mu | \Phi |^2 + \int_{0}^{ | \Phi |^2 } F(U) \upd U.
  \label{pressure}
\end{equation}
In particular, assuming the dependence given in Eq. \ref{index}, the integral term in Eq. \ref{pressure} can be written as  
$ \sum_{q}(-1)^{q+1}\frac{ f_{2 q}}{q + 1}\vert\Phi\vert^{2 (q + 1)}$.

%********************** fig 1  ******************************
\begin{figure}[htbp]
{\centering \resizebox*{1\columnwidth}{!}{\includegraphics{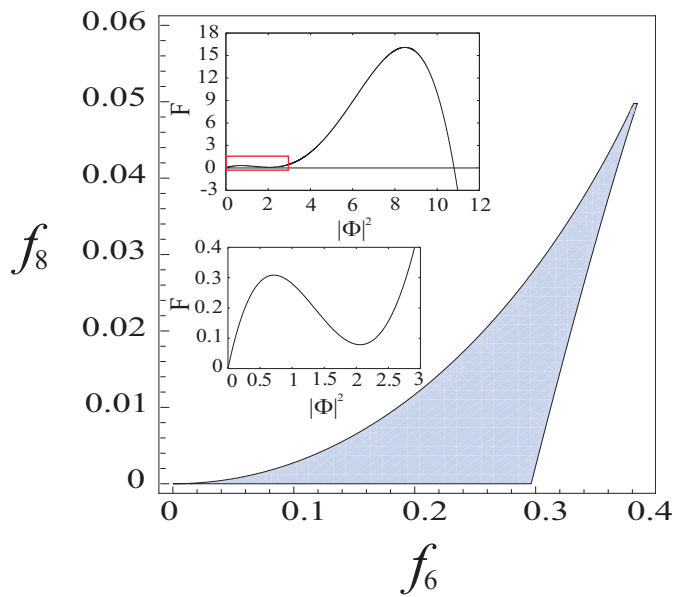}} \par}
\caption{(Colour on-line) Multistability domain given by Eq. \ref{bistability_condition}. Upper inset: nonlinear 
refractive index correction $F$ as a function of the peak intensity of the light beam $|\Phi|^2$. 
Lower inset: detail of the region displayed within the red box depicted in the upper inset.}
\label{fig1}
\end{figure}
%***************************************************************************

In the case of a high-power solution with a flat-top profile of radius $R$ , 
calling $A=\Phi(0)$ the amplitude of the large and homogeneous central region, 
Eq. \ref{NLSE} implies 
\begin{equation}
\mu=-F(A^2),
\label{mu}
\end{equation}
where we choose the arbitrary phase of the solution such that $A$ is positive real.
On the other hand, as shown in Refs. \cite{fermionic,dripping}, any such flat-top solitons 
obey the Young-Laplace (YL) equation\cite{landau}, $p_c = 2\sigma / R$, 
where $R$ is the radius of the droplet,  $p_c=p(0)$ is the central pressure 
and the effective surface tension can be computed as 
$\sigma=-R^{-1}\int_R^\infty r p(r) dr$. In the large $R$ limit, $p_c=0$, 
the gradient term in Eq. \ref{pressure} can be neglected close to the origin, and using Eq. \ref{mu} we get
 \begin{equation}
 \int_{0}^{ A_\infty^2 } F(U)\upd U-A_\infty^2 F(A_\infty^2)=0,
\end{equation}
being $\mu_{\infty}$ and $A_{\infty}$ the asymptotic values corresponding
to the $R \to \infty$ droplet.

%********************** fig 2 *******************************
\begin{figure}[htbp]
{\centering \resizebox*{1\columnwidth}{!}{\includegraphics{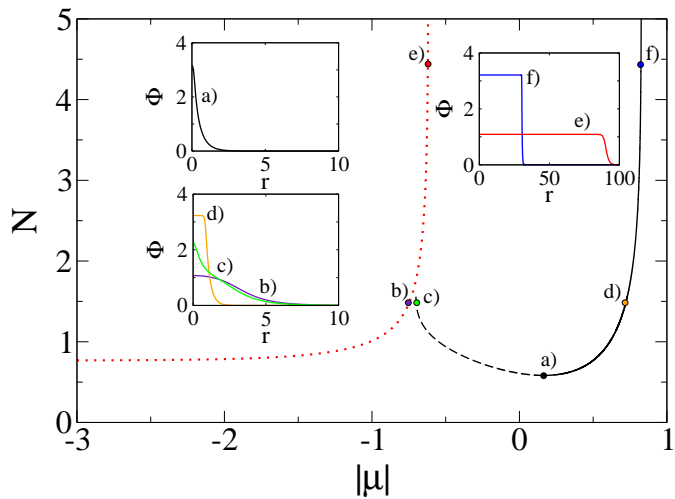}} \par}
\caption{(Colour on-line) Optical power $N$ vs. propagation constant $|\mu|$ for 
the localized nodeless stationary states of the system described by Eq. \ref{NLSE}. 
Both axes display logarithmic scales. The red dotted (black solid) line stands for stable ordinary 
solitons (ultrasolitons), while the black dashed line stands for unstable ultrasolitons. 
Labeled points on the main curves refer to the eigenstates displayed within the insets.
Insets: radial profiles of several eigenstates of the system, whose propagation constants 
are $\mu_a=-1.46=$, $\mu_b=-0.18$, $\mu_c=-0.20$, $\mu_d=-5.26$, $\mu_e=-0.24$ and 
$\mu_f=-6.67$, respectively. Notice that all solitons depicted within each inset feature 
the same power $N$, being this a proof of the existence of OSM.}
\label{fig2}
\end{figure}
%***************************************************************************
 
In particular, for media whose nonlinear response is described by Eq. \ref{index}, 
including terms up to $f_8$, we get the condition
 \begin{equation}
 \frac{1}{2}- \frac{2}{3}U+f_6 \frac{3}{4} U^2-f_8\frac{4}{5} U^3=0,
\label{topflat_condition}
\end{equation}
where $U=A_\infty^2$. This cubic equation, giving $f_6$ and $f_8$,  
can have either one or three real roots. The latter case will eventually correspond to OSM. 
After a long but straightforward algebra, we get the following necessary and sufficient condition 
on $f_6$ and $f_8$ for  Eq. \ref{topflat_condition} to have three real solutions: 
\begin{equation}
18225 f_6^3-5400 f_6^2-77760 f_6 f_8+20480 f_ 8+93312 f_8^2<0.
\label{bistability_condition}
\end{equation}
In Fig. \ref{fig1}, we plot the region of the $(f_6,f_8)$ plane that satisfies such a condition, 
assuming $f_6,f_8>0$. In particular, for $f_6>0.38$ or $f_8>0.05$, condition \ref{bistability_condition}
is not fulfilled and only one solution can be found. 
Conversely, for pairs $(f_6,f_8)$ lying within the shaded region in Fig. \ref{fig1}, 
three different branches of solitons can be found in principle, whose amplitude 
$A_\infty$ for $R\to\infty$ can be obtained from Eq. \ref{topflat_condition}.
In the case of oxygen (air), that was examined in Ref. \cite{fermionic}, $f_6=2.8$ 
and $f_8=3.9$ ($f_6=11.2$ and $f_8=34.1$). 
These values fall out of the shaded region in Fig. \ref{fig1}, 
and in fact only one branch of solitons was found in  Ref. \cite{fermionic}.

Hereafter, we will fix the values $f_6=0.3$ and $f_8=0.02$, that lie in the OSM domain, 
although we have verified that similar results can be obtained for different choices satisfying 
Eq. \ref{bistability_condition}. The nonlinear refractive index dependence on the intensity 
of the input beam acquires then a double-hump structure (see insets in Fig. \ref{fig1}). 
In fact, the emergence of OSM can be heuristically related to the appearance of the two maxima 
of the refractive index, suggesting the possible existence of families of fundamental solitons with limiting intensities close 
to the values corresponding to the two local maxima of $F$. However, the double-hump structure alone 
does not guarantee the emergence of OSM, as we have checked by finding values of ($f_6,f_8$) not satisfying
 the multistability condition \ref{bistability_condition}, that nevertheless lead to a similar double-peak structure for $F$.

%********************** fig 3  ******************************
\begin{figure}[htbp]
{\centering \resizebox*{1\columnwidth}{!}{\includegraphics{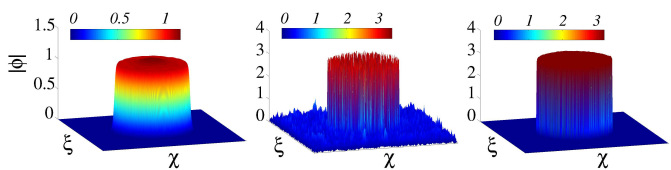}} \par}
\caption{(Colour on-line) 3D pseudocolour plots of the amplitude of three flat-topped beams 
modeled by Eq. \ref{ft_propagation} with amplitudes $\Lambda=A_\infty^o$ (left),  
$A_\infty^g$ (middle), $A_\infty^u$ (right), after a propagation distance of 
$\eta=2000,100$ and $2000$, respectively. They have been initially perturbed with a 
$5\%$ random noise. We see that both beams featuring $A_\infty^o$ and $A_\infty^u$ 
remain stable while the one having $A_\infty^g$ undergoes filamentation. 
The spatial scales spanned are $(\xi,\chi)\in[-100,100]$.}
\label{fig3}
\end{figure}
%***************************************************************************

With the choice $f_6=0.3$, $f_8=0.02$, we calculate the three roots of Eq. \ref{bistability_condition}, 
$A_\infty^o=1.087$, $A_\infty^g=1.601$, $A_\infty^u=3.212$, together with their corresponding 
values for the propagation constant, as given by Eq. \ref{mu},  $\mu_\infty^o=-0.241$, 
$\mu_\infty^g=-0.182$, $\mu_\infty^u=-6.72$. The superscripts $o$, $g$, $u$ refer to 
the names that will be given to the three branches of solitons corresponding to these limiting values, 
namely \emph{ordinary} solitons, \emph{ghost} solitons and \emph{ultrasolitons}, respectively. 

\section{Localized stationary solutions. Ultrasolitons}
We have solved Eq. \ref{NLSE} numerically, and found the localized stationary states 
shown in Fig. \ref{fig2}. We obtain two different branches of solutions that cannot be connected 
with each other, i.e., there are no bifurcations in the eigenstates' structure. 
The red dotted line stands for the \emph{ordinary} soliton branch, i.e., the branch of 
solutions similar to those reported in \cite{fermionic}, whose lower (upper) limit corresponds 
to the critical power \cite{marburger} for self-focusing ($A_\infty^o$ plane wave). 
The presence of HOKE nonlinearities gives rise to a new family of solutions, 
represented by black lines in Fig. \ref{fig2}, whose upper limit corresponds to the 
$A_\infty^u$ plane wave. Their lower limit does not correspond to the Kerr limit,
which means that their existence cannot be explained by a balance between diffraction 
and the leading Kerr nonlinearity driven by $f_2$, but rather as an interplay between competing HOKE nonlinearities. 
To our best knowledge, such solutions do not have counterparts in any other nonlinear optical system 
ruled by local intensity-dependent nonlinearities because they exist over a certain intensity threshold and feature both amplitudes and 
propagation constants higher (in absolute value) than those of the ordinary ensemble. In other words, these solitons belong to 
a completely different nonlinear regime as compared with that of the ordinary branch.
For all the previous reasons, we have called them \emph{ultrasolitons}. 
On the other hand, according to the Vakhitov-Kolokolov (VK) criterium \cite{VK}  supported by systematic simulations of propagation,
we find that the eigenstates represented by the solid (dashed) line are dynamically stable (unstable).

%********************** fig 4  ******************************
\begin{figure}[htbp]
{\centering \resizebox*{1\columnwidth}{!}{\includegraphics{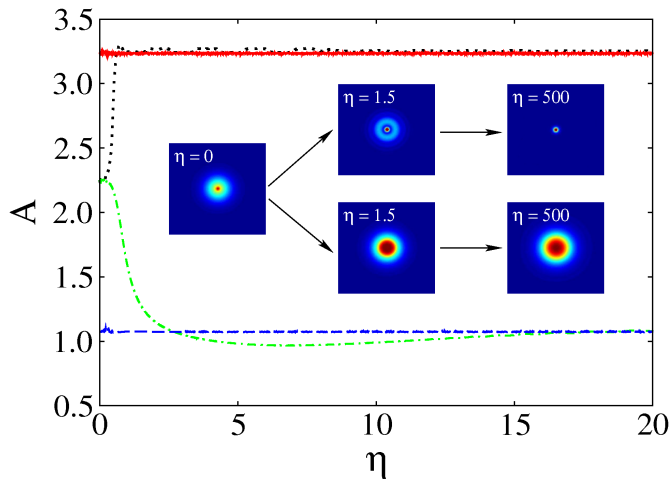}} \par}
\caption{(Colour on-line) Evolution of the peak amplitude $A=\Phi(0,0)$ of the eigenstates 
b) (blue dashed line), c) (green dashed-dotted line) and d) (red solid line) displayed in Fig. \ref{fig2}.
We observe that both b) and d) fields are stable against small perturbations, while the eigenmode 
c) becomes unstable, decaying to the lower branch through a soliton switching mechanism. 
Alternatively, the unstable mode c) can excite a stable ultrasoliton (black dotted line) by adding 
an initial wavefront curvature to the optical field. The upper (lower) row of inner snapshots show 
several pseudocolour amplitude plots of the optical field during the soliton switching procedure, 
where an ultrasoliton (ordinary soliton) is excited from the unstable field c). 
The square window displayed in the snapshots has a width $\omega_{\xi,\chi}=25$.}
\label{fig4}
\end{figure}
%***************************************************************************

The eigenstates b), c) and d) , that are represented in the lower-left inset of Fig. \ref{fig2}, 
feature the same optical power but different propagation constants and radial profiles.  
Analogously, the flat-top eigenstates of insets e) and f) also correspond to the same power, 
thus demonstrating OSM even in the high power regime.
Both cases demonstrate the emergence of OSM, generalizing to our multibranch 
situation the definition given in \cite{kaplan} for a single continuous branch. 

On the other hand, the solution displayed in inset a) of Fig. \ref{fig2} is the first stable 
ultrasoliton, and corresponds to a local minimum of the ultrasoliton power curve depicted 
in Fig. \ref{fig2},  being this a trace of the uniqueness of this solution. Very remarkably, 
this soliton features a \emph{subcritical} power ($N=3.83$ in dimensionless units), 
below the ordinary collapse threshold ($N=5.85$). 
To our best knowledge, this is the first example of a stable subcritical soliton 
in an optical system with an instantaneous nonlinear response. Moreover, this minimum 
power state turns out to have the lowest possible central intensity ($I=9.98$) 
among the stable ultrasolitons, although we have found unstable ultrasolitons 
(corresponding to the dashed curve in Fig. \ref{fig2}) starting from central intensities as small as $I=4.88$.

In Fig. \ref{fig2} we have only plotted two branches of solutions, the ordinary and the ultra- solitons. 
In fact, we have not been able to find numerically the third family of solitary waves linked to the plane 
wave with amplitude $A_\infty^g$ that we have  theoretically predicted above. 
We can understand its non-existence by using three different arguments. 

First,  we recall that, according 
to Refs.  \cite{dripping,fermionic}, for large but 
finite flat-topped solutions the central pressure $p_c$ does not vanish, being compensated 
by the surface tension $\sigma$ like in an ordinary liquid. The actual values of $\sigma$ for 
the three branches can be computed as
\begin{equation}
  \sigma = \frac{1}{\sqrt{2}} \int^{A_{\infty}}_0 \left( -
  \mu_{\infty} A^2 -  \sum_{q = 1}^4 (-1)^{q+1} \frac{f_{2 q} A^{2 (q + 1)}}{q +
  1} \right)^{\frac{1}{2}}\upd A.
\end{equation}

We find: $\sigma^o=0.0948$, $\sigma^g=0.0484 + 0.120 i$ and $\sigma^u=7.21$. 
The fact that for the \emph{ghost} family the surface tension $\sigma^g$ is not real reflects 
the impossibility of equilibrating the inner pressure, which is a real magnitude, 
thus forbidding the existence of finite \emph{ghost} solitons. In other words, 
the ghost solitons would not fulfill the YL equilibrium condition. 

Second, we have applied to the plane wave solutions of our system a linear stability analysis 
similar to that of Refs. \cite{bespalov,filamentCQ}. After a straightforward study, 
we have found that the solutions with $A_\infty^o, A_\infty^u$ are linearly stable, i.e., 
they do not undergo modulational instability under small perturbations, 
while the $A_\infty^g$ lies within an instability window. 

Third, we have studied numerically the propagation of three flat-top beams 
belonging to the high power regime of each of the three branches. 
The initial condition of our simulations is modeled by the following function
\begin{equation}\label{ft_propagation}
\phi=\Lambda(0.25\{[1+\tanh(r+\omega)][1-\tanh(r-\omega)]\}),
\end{equation}
where $\Lambda=$ $A_\infty^o$,  $A_\infty^g$, $A_\infty^u$ is the amplitude of 
the beam envelope and $\omega=50$ is the mean radius. 
We perturb the initial beam profiles with a $5\%$ random noise in order to stimulate the onset of instability. 
The final states arising from the propagation of such initial conditions are displayed in Fig. \ref{fig3}. 
In the left (right) picture, we show the surface amplitude plot of the flat-top beam having an initial 
amplitude $\Lambda=A_\infty^o$ ($A_\infty^u$), at a propagation distance $\eta=2000$. 
We see that such beam is stable, as it quickly couples to a stable eigenstate of the low-intensity 
(high-intensity) branch of ordinary solitons (ultrasolitons) discussed above. 
In the middle picture, we show the outcome of the propagation of an initial beam having 
$\Lambda=A_\infty^g$, at a propagation distance $\eta=100$. 
In this case, we observe how the spatial beam profile has been destabilized 
by the growth of the perturbations, yielding to multiple filamentation like in the 
Cubic-Quintic model\cite{filamentCQ}. The arising filaments correspond to perturbed quasi-stationary ultrasolitons. 
In this context, the onset of filamentation can be considered as an additional trace of the non-existence of the flat-top soliton with $A\approx A_\infty^g$, although it would not be conclusive if taken alone.

However, our present analysis is not sufficient to exclude that triple-stability might 
be found for a different choice of $f_6$ and $f_8$ within the multistability region of Fig. 1. 

\section{Soliton switching}

The existence of OSM, as shown in Fig. \ref{fig2}, suggests the possibility of observing 
transitions between different multistable states. Such beam-reshaping mechanisms 
may lead to \emph{soliton switching} processes with a great potential for all-optical 
communications\cite{kivshar}. In order to study the soliton switching in our system, 
we have simulated the free propagation of three perturbed solitary waves featuring 
the radial profiles b), c) and d) of Fig. \ref{fig2}. The optical power of these beams is 
$N\approx30$. The results of the numerical computations are summarized in Fig. \ref{fig4}, 
plotting the peak amplitude evolution of the eigenstates b) (blue dashed line), 
c) (green dashed-dotted line) and d) (red solid line). 
Even though we have added a $5\%$ random noise to their initial profiles, both fields b) and d) 
remain stable, in agreement with the prediction of the VK criterium applied to our system.
On the other hand, the unstable field c) rapidly decays to a nonlinear mode similar to b), 
even in the absence of external random noise. The conversion efficiency between both modes, 
measured as the ratio between the optical powers of the fields, is above $90\%$. 

We have observed that the unstable fields always decay to the lower branch (\emph{down-switching}), 
while the \emph{up-switching} to the ultrasolitons realm does not occur spontaneously. 
However, we can force such a transition by 
including a focusing quadratic phase term $e^{-i0.01r^2}$, similar 
to that introduced by a thin lens. In this case, the unstable beam d) can be promoted to the 
upper branch (black dotted line in Fig. \ref{fig4}) with a maximum conversion efficiency of 
around $30\%$, while the energy excess is radiated as a low-intensity reservoir resembling 
that generated during the excitation of Townes-like waves in air\cite{townes}. 
The difference between the efficiencies of both processes can be understood by mode-coupling arguments. 
The modal energy transfer becomes more efficient whenever both profiles and phases 
of the corresponding modes match, and the modes b) and c) of Fig. \ref{fig2} have closer profiles 
and propagation constants than the modes c) and d). Few stages 
of the soliton switching process are highlighted in the insets of Fig. \ref{fig4}.

\section{Conclusions}
 
We have shown that an optical system involving competing higher-order 
Kerr nonlinearities can support the existence of power multistability in the absence 
of any external potential, yielding a new class of solitary waves, called \emph{ultrasolitons}, 
that may exhibit powers below the ordinary collapse threshold. 
We have also proposed a mechanism of \emph{soliton switching} for inducing transitions between 
different multistable nonlinear waves of the system, that could have potential applications 
in ultrafast optical circuits intended for all-optical communications\cite{kivshar}. 
We hope that these results will contribute to stimulate the quest 
for a clarification of the mechanism of ultrashort pulses filamentation.

\acknowledgments
D. N. acknowledges support from the MICINN, Spain, through the FCCI. ACI-PROMOCIONA project (ACI2009-1008). 
D. T. thanks the InterTech group at the Universidad Polit\'ecnica de Valencia for hospitality 
during a stay supported by the Government of Spain (PR2010-0415).
H. M.  thanks MPQ at Garching for hospitality during his stay supported by "Salvador de Madariaga" program.

\end{document}